\documentclass[10pt,twoside,titlepage]{article}
\usepackage{amsmath,amssymb}
\usepackage{amsthm}
\usepackage{url}
\usepackage{graphicx}
\usepackage{booktabs}
\usepackage{setspace}
\usepackage{hyperref}
\begin{document}

\title{Extending the Oracle Approximating Shrinkage to General Distributions}
\author{Michael J. Lewis\\
\small Independent Researcher}
\date{\copyright\ October 6, 2026}

\maketitle

\begin{abstract}
The Oracle Approximating Shrinkage (OAS) estimator provides a closed-form, data-driven
shrinkage intensity for regularizing sample covariance matrices toward a scaled identity
target, but its derivation assumes Gaussian data.
In practice, financial returns, neural signals, and many other high-dimensional datasets
exhibit heavy tails and non-zero kurtosis, violating this assumption.
We extend OAS to any distribution with finite fourth moments by incorporating a correction
term $\alpha$ built from the fourth-order joint cumulants of the data.
The correction vanishes under Gaussianity, recovering the OAS formula as a special
case; it admits a consistent $O(np)$ sample estimator and satisfies $\rho\in[0,1]$ exactly
for both the population cumulant and its sample analogue.
Simulations and an equity-returns study confirm that the correction closes the gap between
OAS and the Ledoit--Wolf estimator under non-Gaussian data, with OAS-Extended
performing comparably across all tested settings.
The contribution is a principled correction for OAS practitioners facing heavy-tailed data;
structurally, it is distinct from prior work extending Ledoit--Wolf to elliptical
distributions, which optimizes against a fixed oracle target and yields a different
closed-form formula.

\vfill


\medskip


\noindent\textbf{AMS Classification (MSC 2020):} 62H12 (Primary); 62F10, 91G10 (Secondary)
\end{abstract}

\clearpage

\section{Introduction}\label{sec-intro}

Estimating a covariance matrix from data is a fundamental task across a wide range of
disciplines, including portfolio optimization \cite{LW2004}, signal processing \cite{OAS},
and neuroscience.
In the classical low-dimensional regime, the sample covariance matrix $\hat{S}$ is a
consistent and unbiased estimator of the true covariance $\Sigma$.
However, in the modern high-dimensional setting where the number of variables $p$ is
comparable to or exceeds the number of observations $n$, $\hat{S}$ becomes poorly
conditioned or singular, and its eigenvalues are known to be severely dispersed relative
to those of $\Sigma$.

Shrinkage estimators address this by forming a convex combination
\begin{equation*}
\hat{\Sigma} = (1 - \rho)\hat{S} + \rho \hat{F},
\end{equation*}
where $\hat{F}$ is a well-conditioned target matrix and $\rho \in [0,1]$ is a shrinkage
intensity.
The celebrated Ledoit--Wolf (LW) estimator \cite{LW2004} chooses $\rho$ to minimize the
expected Frobenius loss $E[\lVert \hat{\Sigma} - \Sigma \rVert_F^2]$ against the oracle
target $\hat{F} = (\operatorname{Tr}(\Sigma)/p)\,I$.
The Oracle Approximating Shrinkage (OAS) estimator of Chen et al.\ \cite{OAS} takes a
different approach: it uses the data-dependent target
$\hat{F} = (\operatorname{Tr}(\hat{S})/p)\,I$ and derives $\rho$ via a fixed-point
iteration that converges to a closed-form expression.
This distinction --- random versus fixed target --- leads to structurally different
formulas and different finite-sample behavior, with OAS typically outperforming LW
at small $n/p$ ratios under Gaussian data.

Both estimators were originally derived under the assumption that the data are drawn from
a multivariate Gaussian distribution.
In many practical settings this assumption fails.
Financial asset returns are well-documented to exhibit excess kurtosis and heavy tails.
More broadly, any distribution in the elliptical family --- including the multivariate
Student-$t$ --- introduces fourth-order cumulant structure that the Gaussian-derived
formulas ignore.
Under such heavy-tailed distributions, applying the OAS formula amounts to a misspecified
estimator, and the resulting shrinkage intensity is suboptimal.

The key insight is that the Gaussian derivation of OAS depends on two expectations ---
$E[\operatorname{Tr}(\hat{S}^2)]$ and $E[\operatorname{Tr}^2(\hat{S})]$ --- each of which
acquires an additional term $\alpha/n$ under a general distribution, where
$\alpha = \sum_{i,j} \kappa(X_i, X_i, X_j, X_j)$ aggregates the fourth-order joint
cumulants of the data.
Incorporating this term propagates through the fixed-point iteration and yields a
corrected closed-form shrinkage intensity that reduces to the OAS formula when
$\alpha = 0$.
We also provide a consistent sample estimator $\hat{\alpha}$ and prove that the corrected
formula preserves $\rho \in [0,1]$ for both the population cumulant and the sample
estimator, so an OAS practitioner facing non-Gaussian data can adopt the extended
formula without changing frameworks.

The remainder of the paper is organized as follows.
Section~\ref{sec-related} discusses related work, with particular attention to the
distinction between the present result and prior extensions of LW to elliptical
distributions.
Section~\ref{sec-formulation} derives the corrected oracle shrinkage intensity.
Section~\ref{sec-extension} derives OAS-Extended and its closed form.
Section~\ref{sec-simulation} presents a Monte Carlo simulation study comparing
OAS-Extended and OAS under Gaussian and Student-$t$ data.
Section~\ref{sec-financial} applies all estimators to a rolling-window study on
daily returns of $p=50$ large-cap US equities.
Section~\ref{sec-conclusion} concludes.
The proof that $\rho \in [0,1]$ is given in the Appendix.

\section{Related Work}\label{sec-related}

The closest prior work to this paper is the line of research by Ollila and collaborators
on extending Ledoit--Wolf (LW) shrinkage \cite{LW2004} to elliptical distributions
\cite{Ollila2017, OllilaRaninen2019, Ollila2023}.
In those works, the oracle shrinkage intensity minimizes
$E\left[\lVert (1-\rho)\hat{S} + \rho\mu I - \Sigma \rVert_F^2\right]$
where $\mu = \operatorname{Tr}(\Sigma)/p$ is a \emph{fixed} oracle scalar, yielding a
closed-form expression parameterized by the scalar elliptical kurtosis $\kappa$ and the
sphericity $\gamma = p\operatorname{Tr}(\Sigma^2)/\operatorname{Tr}^2(\Sigma)$.

The OAS framework of \cite{OAS}, which the present work extends, differs from LW in a
structurally important way: the shrinkage target is
$\hat{F} = (\operatorname{Tr}(\hat{S})/p)\,I$, a \emph{random, data-dependent} quantity
rather than the oracle $\mu I$.
This changes the optimization problem and produces a different formula via a fixed-point
iteration; the two approaches are not equivalent even in the Gaussian case.
Consequently, the elliptical-distribution results of \cite{Ollila2017, OllilaRaninen2019}
do not imply an extension of OAS, and to the best of our knowledge the present work is
the first to provide one.

Beyond the structural distinction, our result is also more general in distributional scope.
The Ollila line of work restricts to elliptical distributions, for which the fourth-order
cumulant collapses to a single scalar parameter $\kappa$ via
$\kappa(X_i, X_i, X_j, X_j) = \kappa\bigl(\Sigma_{ii}\Sigma_{jj} + 2\Sigma_{ij}^2\bigr)$.
Our result holds for any distribution with finite fourth moments, where the aggregate cumulant
$\alpha = \sum_{i,j} \kappa(X_i, X_i, X_j, X_j)$ need not have this constrained form.
Elliptical distributions are thus a special case of our framework.

\section{Formulation}\label{sec-formulation}

\paragraph{Setup and notation.}
Let $X^1, \ldots, X^n \in \mathbb{R}^p$ be independent and identically distributed draws
from a centered distribution with covariance matrix $\Sigma$, i.e.\ $E[X^k] = 0$ and
$E[X^k (X^k)^T] = \Sigma$.
We assume the distribution has finite fourth moments.
The index $X_i^k$ denotes the $i$-th component of the $k$-th sample.
The sample covariance matrix is
\begin{equation*}
\hat{S} = \frac{1}{n} \sum_{k=1}^n X^k (X^k)^T,
\end{equation*}
and the shrinkage target is the scaled identity $\hat{F} = \frac{\operatorname{Tr}(\hat{S})}{p} I$,
whose scale is chosen to match the average eigenvalue of $\hat{S}$.
Throughout, $p$ denotes the dimension and $n$ the number of observations.
All expectations are taken over the distribution of the samples.

The setup, notation, and following equations follow \cite{OAS} and serve as our starting point.
\begin{equation}
\rho_O = \frac{ E \left[ \operatorname{Tr} \left( \left( \Sigma - \hat{S} \right) \left( \hat{F} - \hat{S} \right) \right) \right] }{ E \left[ \lVert \hat{S} - \hat{F} \rVert_F^2 \right] }
\end{equation}
\begin{multline}
E \left[ \operatorname{Tr} \left( \left( \Sigma - \hat{S} \right) \left( \hat{F} - \hat{S} \right) \right) \right] =
\frac{\operatorname{Tr}( \Sigma )}{p} E \left[ \operatorname{Tr}(\hat{S})\right]
- \frac{ E \left[ \operatorname{Tr}^2(\hat{S})\right] }{ p } \\
- E \left[ \operatorname{Tr}(\Sigma \hat{S})\right]
+ E \left[ \operatorname{Tr}(\hat{S}^2)\right]
\end{multline}
\begin{equation}
E \left[ \lVert \hat{S} - \hat{F} \rVert_F^2 \right] =
E \left[ \operatorname{Tr}(\hat{S}^2)\right]
- \frac{ E \left[ \operatorname{Tr}^2(\hat{S})\right] }{ p }
\end{equation}
\begin{equation}
E \left[ \operatorname{Tr}(\hat{S})\right] = \operatorname{Tr}(\Sigma)
\end{equation}
Here we depart from \cite{OAS}: rather than assuming Gaussianity, we allow the
distribution to be arbitrary with finite fourth moments.
Observe that
\begin{equation*}
\begin{aligned}
E \left[ \operatorname{Tr}(\hat{S}^2)\right] &= E \left[ \lVert \hat{S} \rVert_F^2 \right] \\
&= \sum_{i,j} \frac{1}{n^2} \sum_{k,l} E \left[ X_i^k X_j^k X_i^l X_j^l \right] \\
&= \sum_{i,j} \frac{1}{n^2} \left( \sum_{k \neq l} E \left[ X_i^k X_j^k X_i^l X_j^l \right] + \sum_{k = l} E \left[ X_i^k X_j^k X_i^l X_j^l \right] \right) \\
&= \frac{n-1}{n} \sum_{i,j} \Sigma^2_{i,j} + \frac{1}{n^2} \sum_{i,j} \sum_k E \left[ (X_i^k)^2 (X_j^k)^2 \right] \\
&= \frac{n-1}{n} \operatorname{Tr}(\Sigma^2) + \frac{1}{n} \sum_{i,j} E \left[ X_i^2 X_j^2 \right] \\
\end{aligned}
\end{equation*}
Similarly,
\begin{equation*}
\begin{aligned}
E \left[ \operatorname{Tr}^2(\hat{S})\right] &= E \left[ \sum_{i,j} \hat{S}_{ii} \hat{S}_{jj} \right] \\
&= \sum_{i,j} \frac{1}{n^2} \sum_{k,l} E \left[  (X_i^k)^2  (X_j^l)^2  \right] \\
&= \sum_{i,j} \frac{1}{n^2} \left( \sum_{k \neq l} E \left[  (X_i^k)^2  (X_j^l)^2  \right] + \sum_{k = l} E \left[  (X_i^k)^2  (X_j^l)^2  \right] \right) \\
&= \frac{n-1}{n} \sum_{i,j} \Sigma_{ii} \Sigma_{jj} + \frac{1}{n^2} \sum_{i,j} \sum_k E \left[ (X_i^k)^2  (X_j^k)^2 \right] \\
&= \frac{n-1}{n} \operatorname{Tr}^2(\Sigma) + \frac{1}{n} \sum_{i,j} E \left[ X_i^2  X_j^2 \right] \\
\end{aligned}
\end{equation*}
The term $E \left[ X_i^2  X_j^2 \right]$ appears in both expansions.
For a centered distribution with finite fourth moments,
\begin{equation}
E \left[ X_i^2  X_j^2 \right] = \Sigma_{ii} \Sigma_{jj} + 2 \Sigma_{ij}^2 + \kappa( X_i, X_i, X_j, X_j )
\end{equation}
where $\kappa( X_i, X_i, X_j, X_j )$ represents the fourth-order joint cumulant.
This cumulant vanishes identically under Gaussianity.
Thus,
\begin{equation}
\label{fourth_moment_eqn}
\sum_{i,j} E \left[ X_i^2  X_j^2 \right] = \operatorname{Tr}^2(\Sigma) + 2 \operatorname{Tr}(\Sigma^2) + \sum_{i,j} \kappa( X_i, X_i, X_j, X_j )
\end{equation}
Let $\alpha =\sum_{i,j} \kappa( X_i, X_i, X_j, X_j )$.

Substituting \eqref{fourth_moment_eqn} into the expansions above,
\begin{equation}
E \left[ \operatorname{Tr}(\hat{S}^2)\right] = \frac{n+1}{n} \operatorname{Tr}(\Sigma^2) + \frac{1}{n} \operatorname{Tr}^2(\Sigma) + \frac{\alpha}{n}
\end{equation}
\begin{equation}
E \left[ \operatorname{Tr}^2(\hat{S})\right] = \operatorname{Tr}^2(\Sigma) + \frac{2}{n} \operatorname{Tr}(\Sigma^2) + \frac{\alpha}{n}
\end{equation}
which match the Gaussian results of \cite{OAS} up to the additional $\frac{\alpha}{n}$ term.
Substituting into the oracle formula and simplifying yields the corrected oracle intensity:
\begin{equation}
\label{opt_rho}
\rho_O = \frac{ (1-2/p) \operatorname{Tr}(\Sigma^2) + \operatorname{Tr}^2(\Sigma) + \alpha (1 - 1/p) }{ (n+1-2/p) \operatorname{Tr}(\Sigma^2) + (1 - n/p) \operatorname{Tr}^2(\Sigma) + \alpha (1 - 1/p) }
\end{equation}

\section{Extending OAS}\label{sec-extension}
We now apply the OAS fixed-point framework of \cite{OAS} to the corrected oracle formula \eqref{opt_rho}.
As in \cite{OAS}, we have
\begin{equation}
\label{rho_eval}
\hat{\rho}_{j+1} = \frac{ (1-2/p) \operatorname{Tr}(\hat{\Sigma}_j \hat{S} ) + \operatorname{Tr}^2( \hat{\Sigma}_j ) + \alpha (1 - 1/p) }{ (n+1-2/p) \operatorname{Tr}( \hat{\Sigma}_j \hat{S} ) + (1 - n/p) \operatorname{Tr}^2( \hat{\Sigma}_j ) + \alpha (1 - 1/p) } \\
\end{equation}
\begin{equation}
\label{sigma_eval}
\hat{\Sigma}_{j+1} = (1 - \hat{\rho}_{j+1}) \hat{S} + \hat{\rho}_{j+1} \hat{F} \\
\end{equation}
which leverages \eqref{opt_rho}, but replaces $\operatorname{Tr}(\Sigma)$ and $\operatorname{Tr}(\Sigma^2)$ with
$\operatorname{Tr}(\hat{\Sigma}_j)$ and $\operatorname{Tr}(\hat{\Sigma}_j \hat{S})$, respectively.
Let
\begin{equation}
\hat{\phi} = \frac{ \operatorname{Tr}( \hat{S}^2 ) - \operatorname{Tr}^2( \hat{S} ) / p }{ (1-2/p) \operatorname{Tr}( \hat{S}^2 ) + \operatorname{Tr}^2( \hat{S} ) } = \frac{ \operatorname{Tr}( \hat{S}^2 ) - \operatorname{Tr}^2( \hat{S} ) / p }{ M }
\end{equation}
and $\beta = \alpha (1 -  1/p) / M$.
Substituting \eqref{sigma_eval} into \eqref{rho_eval} and simplifying yields
\begin{equation}
\hat{\rho}_{j+1} = \frac{ 1 + \beta - (1 - 2/p) \hat{\phi} \hat{\rho}_j }{ 1 + \beta + n \hat{\phi} - (n+1-2/p) \hat{\phi} \hat{\rho}_j }
\end{equation}
Using the change of variables
\begin{equation}
b_j  = \frac{1}{ 1 + \beta - (n+1-2/p) \hat{\phi} \hat{\rho}_j } \Leftrightarrow \hat{\rho}_j = \frac{ 1 + \beta - \frac{1}{b_j} }{ (n+1-2/p) \hat{\phi} },
\end{equation}
showing that \eqref{rho_eval} is equivalent to
\begin{equation}
b_{j+1} =  \frac{  n \hat{\phi} }{ 1 + \beta - (1-2/p) \hat{\phi} } b_j + \frac{ 1 }{  1 + \beta - (1-2/p) \hat{\phi} }
\end{equation}
Since this is a linear recurrence in $b_j$, its limit is
\begin{equation}
  \lim_{j \to \infty} b_j =
    \begin{cases}
      \infty & \frac{  n \hat{\phi} }{ 1 + \beta - (1-2/p) \hat{\phi} } \geq  1 \\
      \frac{1}{1 + \beta - (n+1-2/p) \hat{\phi} }      & \frac{  n \hat{\phi} }{ 1 + \beta - (1-2/p) \hat{\phi} } < 1 \\
    \end{cases}
\end{equation}
This implies
\begin{equation}
  \lim_{j \to \infty} \hat{\rho}_j =
    \begin{cases}
      \frac{ 1 + \beta }{ (n+1-2/p) \hat{\phi} } & (n+1-2/p) \hat{\phi} \geq 1 + \beta \\
      1                                          & (n+1-2/p) \hat{\phi} < 1 + \beta \\
    \end{cases}
\end{equation}
or equivalently,
\begin{equation}
\label{new_OAS}
\rho_{\text{OAS-Extended}} = \min\left( 1, \frac{ 1 + \beta }{ (n+1-2/p) \hat{\phi} } \right)
\end{equation}

That $1 + \beta \geq 0$ is established in Appendix~\ref{app:proof}, ensuring $\rho_{\text{OAS-Extended}} \geq 0$,
so \eqref{new_OAS} defines a valid shrinkage intensity in $[0,1]$.

It remains to construct a consistent estimator for $\beta$.
Since $M = (1-2/p) \operatorname{Tr}( \hat{S}^2 ) + \operatorname{Tr}^2( \hat{S} )$ is computable from $\hat{S}$, it suffices to estimate $\alpha$.
By its definition and \eqref{fourth_moment_eqn}, the natural estimator is
\begin{equation}
\hat{\alpha} = \left( \sum_{i,j} \frac{1}{n} \sum_k (X_i^k)^2 (X_j^k)^2 \right) - \operatorname{Tr}^2(\hat{S}) - 2 \operatorname{Tr}(\hat{S}^2),
\end{equation}
which converges to the true $\alpha$ by the law of large numbers.
Noting that $\sum_{i,j}\frac{1}{n}\sum_k (X_i^k)^2(X_j^k)^2 = \frac{1}{n}\sum_k \lVert X^k \rVert^4$
and $\operatorname{Tr}(\hat{S}) = \frac{1}{n}\sum_k \lVert X^k \rVert^2$, the estimator simplifies to
\begin{equation}
\hat{\alpha} = \widehat{\operatorname{Var}}\!\left(\lVert X \rVert^2\right) - 2\operatorname{Tr}(\hat{S}^2),
\end{equation}
where $\widehat{\operatorname{Var}}(\lVert X \rVert^2) = \frac{1}{n}\sum_k \lVert X^k \rVert^4 - \operatorname{Tr}^2(\hat{S})$
is the sample variance of the squared norms.
This form has a natural interpretation: $\hat{\alpha}$ measures the excess squared-norm
variance beyond what a Gaussian with covariance $\hat{S}$ would produce
($2\operatorname{Tr}(\hat{S}^2)$ being the Gaussian value), and is computable in $O(np)$
additional time once $\hat{S}$ is formed.
Substituting $\hat{\beta}$ for $\beta$ in \eqref{new_OAS} gives the sample estimator.
Appendix~\ref{app:proof} shows that $1 + \hat{\beta} \geq 0$ holds exactly for
every finite sample, so $\hat{\rho}_{\text{OAS-Extended}} \in [0,1]$ without any clamping.

\section{Simulation Study}\label{sec-simulation}

We evaluate OAS-Extended against four competitors: OAS \cite{OAS},
the Rao-Blackwell Ledoit--Wolf (RBLW) estimator \cite{OAS},
the analytical Ledoit--Wolf (LW) estimator \cite{LW2004}, and the
sample covariance matrix.
All experiments use a $p=120$ dimensional AR(1) covariance structure
$\Sigma_{ij} = 0.5^{|i-j|}$, with $n/p \in \{0.25, 0.5, 1, 2, 5\}$ and 500 independent
trials per setting.
The smallest ratio $n/p = 0.25$ corresponds to $n = 30$ observations, ensuring the
sample is large enough for stable estimation of $\hat{\alpha}$ while remaining well
into the high-dimensional regime where $\hat{S}$ is singular.
Performance is measured by the normalized Frobenius loss
$\lVert \hat{\Sigma} - \Sigma \rVert_F / \lVert \Sigma \rVert_F$.

Three distributions are considered.
The Gaussian case ($\alpha = 0$) serves as a sanity check.
We then draw from a multivariate Student-$t$ with $\nu = 10$ degrees of freedom
(excess kurtosis $= 1$) and $\nu = 5$ (excess kurtosis $= 6$),
both scaled so that $\operatorname{Cov}(X) = \Sigma$ exactly.
Both choices satisfy $\nu > 4$, the requirement for finite fourth moments.
For a multivariate Student-$t$ with $\nu$ degrees of freedom, the elliptical family
formula gives $\alpha = \frac{2}{\nu-4}\bigl[\operatorname{Tr}^2(\Sigma) + 2\operatorname{Tr}(\Sigma^2)\bigr]$,
which grows rapidly as $\nu$ decreases toward 4, explaining why the $\alpha$ correction
is substantially larger under $t(\nu\!=\!5)$ than $t(\nu\!=\!10)$.

Table~\ref{tab:sim} reports mean losses.
Under Gaussian data all four shrinkage estimators perform comparably (confirming no bias
when $\alpha = 0$), with the minor exception of $n/p = 0.25$ where $\hat{\alpha}$
estimation noise marginally inflates OAS-Extended's intensity; this gap closes by
$n/p = 0.5$.
Under non-Gaussian data, four findings stand out.
First, the correction recovers the performance lost by OAS's Gaussian
misspecification: OAS-Extended performs comparably to LW, with differences small at
$n/p = 0.25$ and negligible at $n/p \geq 1$.
Second, both dominate OAS substantially under $t(\nu\!=\!5)$: losses fall
by 46--48\% at $n/p \leq 0.5$, $\sim$36\% at $n/p = 1$, $\sim$24\% at $n/p = 2$, and $\sim$16\% at $n/p = 5$.
Third, RBLW performs comparably to OAS under non-Gaussian data and in some
settings is slightly worse, confirming that neither the Rao-Blackwell construction of
RBLW nor the fixed-point iteration of OAS provides a non-Gaussian correction.
Fourth, the improvement of OAS-Extended over OAS grows with kurtosis and
shrinks as $n/p$ increases, exactly as predicted by the theory.

\begin{figure}[ht]
\centering
\includegraphics[width=\textwidth]{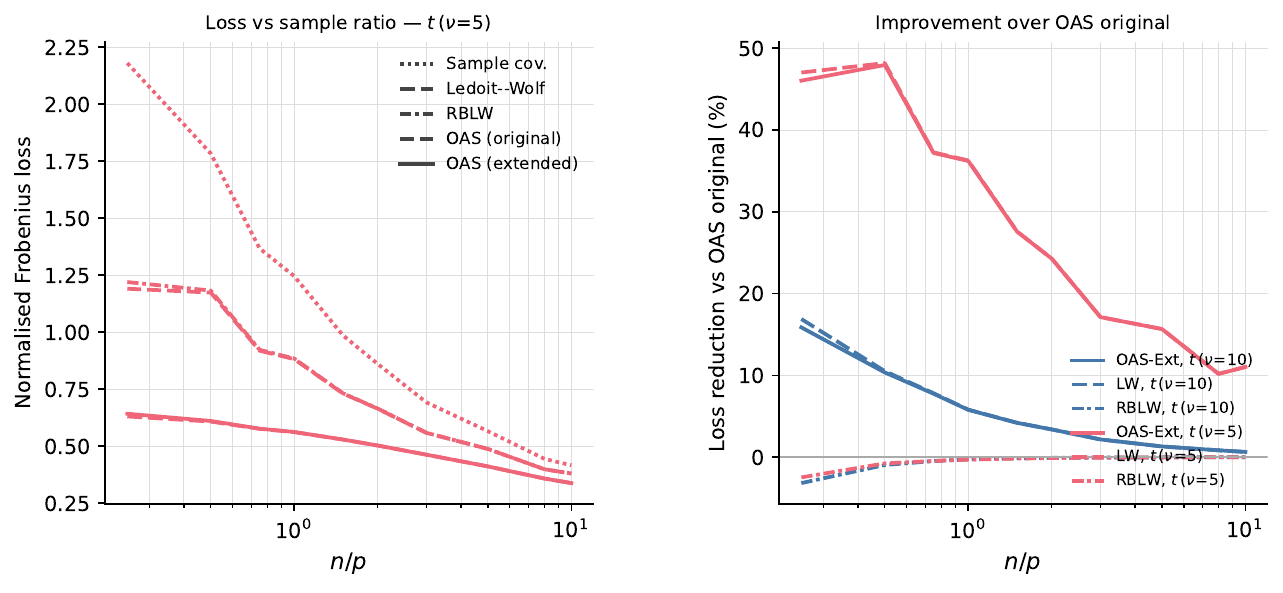}
\caption{Left: normalized Frobenius loss vs $n/p$ for $t(\nu\!=\!5)$ data ($p=120$), all five
estimators (dotted = sample covariance, long-dash = Ledoit--Wolf, dash-dot = RBLW,
dashed = OAS, solid = OAS-Extended). Right: relative loss reduction vs OAS (\%) for OAS-Extended (solid), Ledoit--Wolf (long-dash), and RBLW (dash-dot),
across both non-Gaussian distributions (blue = $t(\nu\!=\!10)$, red = $t(\nu\!=\!5)$),
over the full range $n/p \in [0.25, 10]$.
OAS-Extended performs comparably to LW, both substantially outperforming OAS
and RBLW under heavy-tailed data.}
\label{fig:sim}
\end{figure}

Figure~\ref{fig:sim} shows loss curves and relative improvement over the full range
$n/p \in [0.25, 10]$.

\begin{table}[ht]
\centering
\caption{Mean normalized Frobenius loss ($p=120$, 500 trials). Lower is better.}
\label{tab:sim}
\begin{tabular}{clccccc}
\toprule
$n/p$ & Distribution & Sample cov. & LW & RBLW & OAS & OAS (extended) \\
\midrule
0.25 & Gaussian              & 1.5602 & 0.5884 & 0.5883 & 0.5866 & 0.5935 \\
     & $t$ ($\nu\!=\!10$) & 1.8011 & 0.6046 & 0.7504 & 0.7275 & 0.6120 \\
     & $t$ ($\nu\!=\!5$) & 2.1799 & 0.6308 & 1.2200 & 1.1907 & 0.6426 \\
\midrule
0.5 & Gaussian              & 1.1035 & 0.5486 & 0.5486 & 0.5484 & 0.5492 \\
    & $t$ ($\nu\!=\!10$) & 1.2652 & 0.5680 & 0.6408 & 0.6348 & 0.5690 \\
    & $t$ ($\nu\!=\!5$) & 1.7838 & 0.6083 & 1.1826 & 1.1738 & 0.6110 \\
\midrule
1.0 & Gaussian              & 0.7810 & 0.4908 & 0.4907 & 0.4907 & 0.4908 \\
    & $t$ ($\nu\!=\!10$) & 0.8962 & 0.5176 & 0.5510 & 0.5496 & 0.5177 \\
    & $t$ ($\nu\!=\!5$) & 1.2458 & 0.5621 & 0.8847 & 0.8821 & 0.5627 \\
\midrule
2.0 & Gaussian              & 0.5530 & 0.4156 & 0.4156 & 0.4156 & 0.4156 \\
    & $t$ ($\nu\!=\!10$) & 0.6397 & 0.4494 & 0.4656 & 0.4653 & 0.4495 \\
    & $t$ ($\nu\!=\!5$) & 0.8613 & 0.5033 & 0.6654 & 0.6648 & 0.5034 \\
\midrule
5.0 & Gaussian              & 0.3495 & 0.3056 & 0.3056 & 0.3056 & 0.3056 \\
    & $t$ ($\nu\!=\!10$) & 0.4039 & 0.3403 & 0.3449 & 0.3449 & 0.3403 \\
    & $t$ ($\nu\!=\!5$) & 0.5660 & 0.4116 & 0.4882 & 0.4882 & 0.4116 \\
\bottomrule
\end{tabular}
\end{table}

\clearpage

\section{Empirical Study: Financial Returns}\label{sec-financial}

We complement the Monte Carlo evidence with an application to daily equity returns.
The universe consists of $p = 50$ large-cap US equities drawn from the S\&P~500 index;
we use adjusted daily log-returns from January 2010 to December 2023
($T = 3{,}521$ trading days), downloaded via the \texttt{yfinance} library.
Returns are demeaned within each estimation window.

\paragraph{Rolling-window evaluation.}
We adopt a rolling-window protocol with in-sample window $T_{\mathrm{in}} = 252$ trading
days (approximately one trading year, giving $n/p \approx 5$) and an out-of-sample
horizon of 21 trading days (one calendar month), stepped forward monthly, yielding
$N = 155$ evaluation windows.
At each step we estimate the covariance matrix with each of the five estimators and form
the global minimum variance (GMV) portfolio $\hat{w} \propto \hat{\Sigma}^{-1}\mathbf{1}$.
Performance is measured by the mean annualized realized volatility
$100\% \times \sqrt{252 \cdot \overline{\operatorname{Var}(\hat{w}^{\!\top} r_{\mathrm{out}})}}$.

\begin{figure}[ht]
\centering
\includegraphics[width=\textwidth]{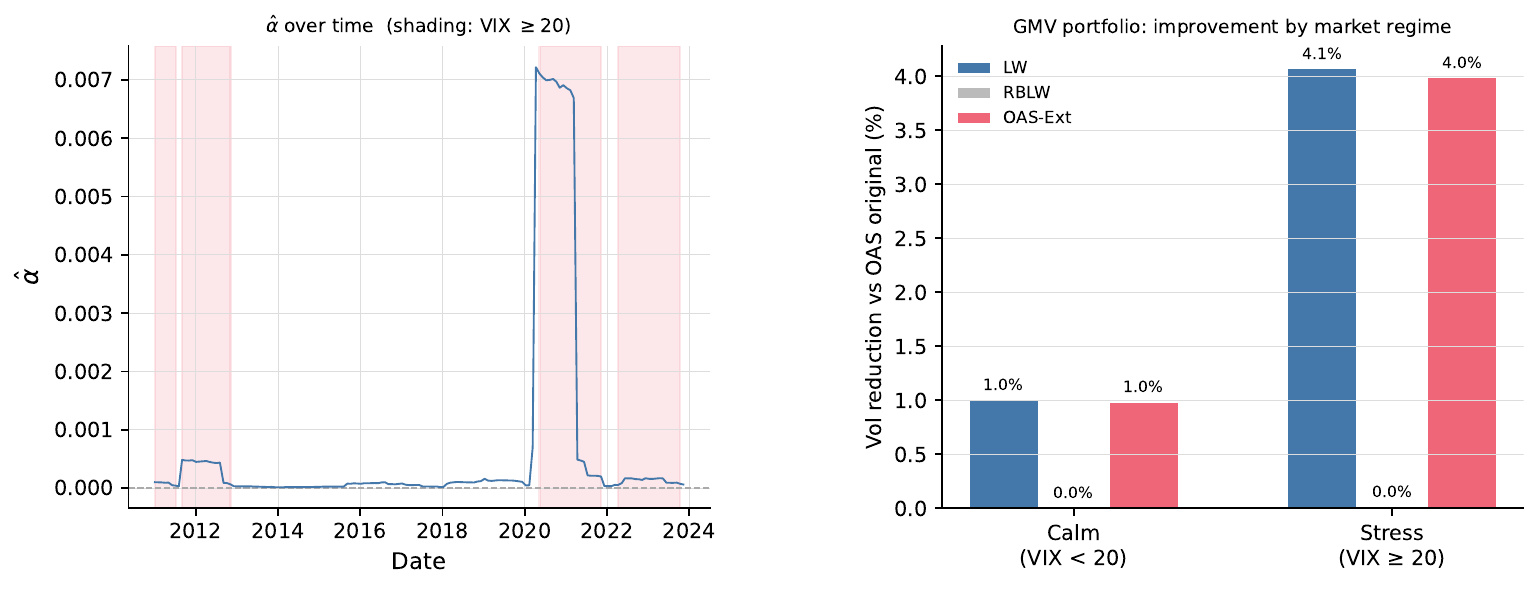}
\caption{Left: rolling-window $\hat{\alpha}$ for $p=50$ large-cap US equities
(January 2010 -- December 2023), with stress periods (mean in-sample VIX $\geq 20$)
shaded in red.
$\hat{\alpha}$ is positive in every window; spikes coincide with the European
sovereign-debt crisis (2011--2012) and the COVID-19 dislocation (2020).
Right: percentage improvement in GMV portfolio realized volatility over OAS,
split by market regime.
OAS-Extended and LW both deliver $\sim 1\%$ vol reduction in calm markets and
$\sim 4\%$ in stress periods, four times larger, while RBLW's improvement over OAS is statistically detectable but economically negligible ($< 0.001\%$ vol
reduction in both regimes).}
\label{fig:financial}
\end{figure}

\paragraph{Non-Gaussianity.}
Figure~\ref{fig:financial} (left) plots $\hat{\alpha}$ over the 155 rolling windows.
The estimate is positive in every window (median $\hat{\alpha} = 9.0 \times 10^{-5}$,
mean $6.5 \times 10^{-4}$), confirming the well-documented heavy-tailed character of
equity returns.
The largest values coincide with periods of elevated market stress; the spike in early 2020
corresponds to the COVID-19 market dislocation.

\paragraph{Portfolio results.}
Table~\ref{tab:financial} reports mean annualized realized volatility for each estimator.
OAS-Extended reduces realized volatility by 2.0\% relative to OAS
(13.28\% vs.\ 13.55\%), approaching LW (13.282\% vs.\ 13.276\%),
consistent with the simulation findings.
RBLW closely tracks OAS (13.55\%), again mirroring the Monte Carlo results.
The improvement of OAS-Extended over OAS is highly significant
($t = 3.83$, $p < 0.001$; bootstrap 95\% CI for the mean variance reduction:
$[2.0, 5.0] \times 10^{-6}$).
LW achieves the same significance ($t = 3.85$, $p < 0.001$).
RBLW also reaches formal significance ($t = 3.88$, $p < 0.001$), but its mean variance
reduction is below $10^{-7}$ --- more than 30 times smaller than OAS-Extended's ---
and the bootstrap 95\% CI rounds to zero at six decimal places.
The result is statistically detectable but economically negligible, confirming that
neither RBLW nor OAS provides a meaningful non-Gaussian correction.

\paragraph{Regime breakdown.}
We classify each window as \emph{stress} if the mean VIX over the in-sample period
exceeds 20 (a standard threshold for elevated implied volatility), and \emph{calm}
otherwise.
Of the 155 windows, 99 are calm and 56 are stress.
Table~\ref{tab:regime} and Figure~\ref{fig:financial} (right) report the improvement
over OAS by regime.
In calm windows the correction yields a 1.0\% vol reduction.
In stress windows the improvement rises to 4.0\%, approximately four times larger.
This concentration of the gain in high-$\hat{\alpha}$ periods is the direct
empirical counterpart of the theory: when heavy-tailed behavior is most pronounced,
the correction to the shrinkage intensity matters most.

\begin{table}[ht]
\centering
\caption{Mean annualized realized GMV portfolio volatility (\%), $p=50$ large-cap
US equities, January 2010 -- December 2023, $N=155$ rolling windows of
$T_{\mathrm{in}}=252$ days. Lower is better.}
\label{tab:financial}
\begin{tabular}{lccccc}
\toprule
& Sample cov. & LW & RBLW & OAS & OAS (extended) \\
\midrule
All windows ($N=155$) & 13.972 & 13.276 & 13.551 & 13.552 & 13.282 \\
\bottomrule
\end{tabular}
\end{table}

\begin{table}[ht]
\centering
\caption{Mean annualized realized GMV portfolio volatility (\%) and percentage vol
reduction relative to OAS, by market regime (VIX threshold = 20).
Parenthesized values are the percentage vol reduction vs OAS.}
\label{tab:regime}
\begin{tabular}{lccccc}
\toprule
Regime & $N$ & OAS & LW & RBLW & OAS (extended) \\
\midrule
Calm (VIX $< 20$)      & 99 & 13.785 & 13.645 ($+1.0\%$) & 13.784 ($+0.0\%$) & 13.649 ($+1.0\%$) \\
Stress (VIX $\geq 20$) & 56 & 13.131 & 12.596 ($+4.1\%$) & 13.129 ($+0.0\%$) & 12.607 ($+4.0\%$) \\
\bottomrule
\end{tabular}
\end{table}

\section{Conclusion}\label{sec-conclusion}

We have derived a closed-form correction to the Oracle Approximating Shrinkage estimator
valid for any distribution with finite fourth moments.
The correction enters through a single scalar $\alpha$ aggregating the fourth-order joint
cumulants of the data; it vanishes under Gaussianity, recovering the OAS formula
as a special case.
A consistent $O(np)$ sample estimator $\hat{\alpha}$ is available in closed form, and the
corrected shrinkage intensity lies in $[0,1]$ exactly for both the population cumulant and
its sample analogue.

Simulations and a rolling-window equity study confirm that the correction closes the gap
between OAS and Ledoit--Wolf under heavy-tailed data, with gains largest at small
$n/p$ and in high-stress market regimes.
The correction itself requires only finite fourth moments and applies to real-world data
regardless of parametric form.
The simulation study is, however, limited to Student-$t$ distributions and an AR(1)
covariance structure, and the financial application covers a single universe of US
large-cap equities; generalization of the empirical findings to other settings remains
to be assessed.
Natural extensions include applying the framework to structured shrinkage targets such as
factor models or banded covariance matrices, and examining the behavior of $\hat{\alpha}$
in very high-dimensional settings where $n/p \ll 1$.

\clearpage

\appendix
\section{Proof That $1 + \beta \geq 0$}
\label{app:proof}
It is clear from \eqref{new_OAS} that $\rho_{\text{OAS-Extended}} \leq 1$.
We show $1 + \beta \geq 0$, establishing the lower bound.
The argument is identical for the population quantity $\beta$ and the sample estimator
$\hat{\beta}$; we write $A$ for either $\Sigma$ or $\hat{S}$, and $V$ for either
$\operatorname{Var}(\lVert X \rVert^2)$ or $\widehat{\operatorname{Var}}(\lVert X \rVert^2)$.

\begin{proof}
Both variance quantities are non-negative: the population variance by definition, and
\[
\widehat{\operatorname{Var}}(\lVert X \rVert^2)
  = \tfrac{1}{n}\sum_k \lVert X^k \rVert^4 - \operatorname{Tr}^2(\hat{S}) \geq 0
\]
by Jensen's inequality on the empirical distribution.
Since $\alpha = V - 2\operatorname{Tr}(A^2)$, we have $\alpha \geq -2\operatorname{Tr}(A^2)$.
Noting $\beta = \alpha(1-1/p)/M$, it suffices to show $M \geq 2(1-1/p)\operatorname{Tr}(A^2)$,
i.e.\ $\operatorname{Tr}^2(A) \geq \operatorname{Tr}(A^2)$.
For any positive semidefinite $A$ with eigenvalues $\lambda_1, \ldots, \lambda_p \geq 0$,
\[
\operatorname{Tr}^2(A) - \operatorname{Tr}(A^2)
= \Bigl(\sum_i \lambda_i\Bigr)^2 - \sum_i \lambda_i^2
= 2\sum_{i < j} \lambda_i \lambda_j \geq 0.
\]
Hence $1 + \beta \geq 0$, so $\rho_{\text{OAS-Extended}} \in [0,1]$ for both the
population cumulant and the sample estimator.
\end{proof}

\section*{AI Use Disclosure}
The author used Claude (Anthropic) to assist with prose editing,
readability improvements, simulation code, and empirical analyses.
All mathematical derivations and intellectual contributions are
solely the author's own work.

\end{document}